\documentclass[3p,times,procedia]{elsarticle}
\usepackage{nupha_ecrc}

\volume{00}

\firstpage{1}

\journalname{Nuclear Physics A}

\runauth{Scott McDonald, Sangyong Jeon, Charles Gale}

\jid{nupha}

\jnltitlelogo{Nuclear Physics A}

\usepackage{amssymb}
\usepackage{amsmath}
\begin{document}

\begin{frontmatter}



%% Instructions from Editor: Please use the following \dochead only in the preprint version (e-print arXiv etc.); 
%% use empty \dochead{} when submitting to Nuclear Physics A!
\dochead{XXVIIth International Conference on Ultrarelativistic Nucleus-Nucleus Collisions\\ (Quark Matter 2018)}
%% e.g. \dochead{17th International Conference on Dynamical Processes in Excited States of Solids}

\title{IP-Glasma Phenomenology Beyond 2D}

\author{Scott McDonald}

\author{Sangyong Jeon}
 
 \author{Charles Gale}
 
 \address{Department of Physics, McGill University, 3600 University
 Street, Montreal,
 QC, H3A 2T8, Canada}

\begin{abstract}
We present a novel formulation of the IP-Glasma initial state model in 3+1D in which the 2+1D boost invariant IP-Glasma is generalized through JIMWLK rapidity evolution of the pre-collision Wilson lines \cite{Schenke:2016ksl}. In order to consistently accommodate a non-boost-invariant system, the initial conditions are generalized to 3+1D. By breaking boost invariance, the 3+1D model no longer trivially satisfies Gauss' law at the initial time, and we now enforce it locally. We compare the time evolution of the chromo-electric and chromo-magnetic fields as well as the transverse and longitudinal pressures in the 3+1D case with the boost invariant result.
\end{abstract}


\end{frontmatter}

% \linenumbers

\section{Introduction}
Heavy ion collisions (HIC's) at the LHC and RHIC are believed to create a hot, dense state of deconfined quarks and gluons known as Quark Gluon Plasma (QGP). Phenomenologically, simulations of HIC's are generally modelled in stages, including an initial stage that describes the first few $\rm{fm}$ after the collision, before the formation of an equilibrated QGP medium. 

IP-Glasma \cite{Schenke:2012wb, Schenke:2012fw} is a 2+1D initial condition model that combines IP-Sat \cite{Kowalski:2003hm} inspired small-$x$ gluon saturation with Classical Yang-Mills (CYM) evolution. It has been extremely successful in describing the transverse dynamics of HIC's when used to initialize hydrodynamic simulations. As a 2+1D initial state model, however, IP-Glasma is unable to describe the longitudinal dynamics of HIC's.

The current work \cite{InPrep} extends the IP-Glasma framework to 3+1D by generalizing the 2+1D initial conditions, using JIMWLK renormalization group equations to determine the rapidity evolution of the small-$x$ gluons, and solving the CYM equations on a 3-dimensional lattice. We study the consequences of generalizing the model to 3+1D by comparing it to the boost invariant case. In particular, we investigate the time evolution of the chromo-electric and chromo-magnetic fields, as well as the longitudinal and transverse pressures. 
\section{Initial Conditions in 2+1D}
Under the assumption of infinite momentum nuclei, it is possible to derive an analytic solution to the initial gauge fields immediately following the collision. The solution relates the strictly transverse pure gauge fields of the pre-collision nuclei to the solution in the forward light cone by matching the fields on the light cone boundary. The pure gauge fields of the pre-collision nuclei, A and B, can be written in terms of the Wilson line,
%==========================================%
\begin{align}
    A^{A(B)}_{i=x,y} = \frac{i}{g}V^{A(B)} \partial_i V^{A(B)\dagger}  && A^{A(B)}_{\eta} = 0 && \hbox{with} && V^{A(B)}(\mathbf{x_\perp})=\prod_{i=1}^{N_y=50}\exp\Big({-ig\frac{\rho^{A(B)}_i(x_\perp)}{\nabla^2-m^2}}\Big)
\end{align}
%==========================================%
where $\rho$ is color charge density, and $m$ is an infrared regulator. The fields in the forward light cone are then given in terms of the fields of nuclei A and B, by 
%==========================================%
\begin{align}\label{init_2D3HOm}
  A_{i}=A^{A}_{i}+A^{B}_{i} &&
A^{\eta}=-\frac{1}{2}E^{\eta}=\frac{ig}{2}[A^{A}_{i},A^{B}_{i}].
\end{align}
%==========================================%
It only remains to determine the transverse electric fields to completely specify the initial condition. This is done using Gauss' law, given by
%==========================================%
\begin{equation}\label{eq:gauss}
    [D_\eta, E^\eta]+[D_i,E^i]=0.
\end{equation}
%==========================================%
 The boost invariance of the system makes Gauss' Law trivial due to vanishing derivatives in $\eta$. The resulting solution is $E^i(\tau=0)=0$. Similarly, the initial transverse magnetic fields, $B_i=F_{j\eta}=\partial_j A_\eta - \partial_\eta A_j -ig[A_j, A_\eta]$ vanish because $A_\eta(\tau=0)=0$ and  $\eta$-derivatives vanish in the boost invariant case. It is important to note that the initial transverse $E$ and $B$ fields only truly vanish under the assumption of infinite momentum.
\section{Initial Conditions in 3+1D}
In recent years, significant progress has been made on 3+1D initial conditions for heavy ion collisions \cite{Schenke:2016ksl,Gelis:2013rba, Romatschke:2006nk, Gelfand:2016yho, Kapusta:2016vhl, Shen:2017bsr}. Most of these efforts have implemented some type of rapidity dependence, whether through JIMWLK evolution \cite{Schenke:2016ksl}, rapidity fluctuations \cite{Gelis:2013rba, Romatschke:2006nk}, the inclusion of dynamical color sources \cite{Gelfand:2016yho}, or the dynamical initialization of hydrodynamics with sources \cite{Shen:2017bsr}. However, all of the CYM studies mentioned have used the 2+1D boost invariant initial conditions for the gauge fields discussed in the previous section, either directly or as a background on which perturbations were introduced. Here, we extend these initial conditions themselves to be able to consistently accommodate a non-boost-invariant setup.

In order to ensure that the energy density vanishes outside of the interaction region, the pre-collision gauge fields of each nuclei must be pure gauge, as they are in 2+1D. In 3+1D, non-zero gradients in the rapidity direction necessitate the inclusion of a non-zero $\eta$-component,
%==========================================%
\begin{align}\label{eq:Aeta}
      A_{i=x,y,\eta}=A^{A}_{i}+A^{B}_{i} && A^{A(B)}_{i=x,y,\eta} = \frac{i}{g}V^{A(B)} \partial_i V^{A(B)\dagger}.
\end{align}
%==========================================%
This is a natural extension of the initial gauge fields from the 2+1D case: in the boost invariant limit, the $\eta$-derivative vanishes and the 2+1D initial condition for the single nucleus gauge fields is recovered. As a consequence, $B_i=F_{j\eta}=\partial_j A_\eta - \partial_\eta A_j -ig[A_j, A_\eta] \neq 0$, at initial time.

In order to completely specify the initial condition, it is also necessary to determine the initial electric fields. The initial longitudinal electric field is unchanged from the 2+1D case, and is given by $E^{\eta}=-ig[A_{A}^{i},A_{B}^{i}]$. At finite energies, however, Gauss' Law, which determines the initial transverse electric fields, is no longer trivial. In fact, Gauss' Law is under-constrained, with one equation for two unknown fields, $E^x$ and $E^y$. We relate them through the following ansatz 
%==========================================%
\begin{equation}
    E^i=[D^i,\phi] 
\end{equation}
%==========================================%
which turns Gauss' Law into the covariant Poisson equation,
%==========================================%
\begin{equation}
    [D_\eta, E^\eta]+[D_i,E^i]=[D_\eta, E^\eta]+[D_i,[D^i,\phi]]=0.
\end{equation}
%==========================================%
This can be solved iteratively through the Jacobi method. Thus, we have a solution to Gauss' Law in the non-boost invariant system, and the transverse electric fields are, in general, non-zero at initial time.

\section{Rapidity Dependence - JIMWLK Evolution}\label{JIMWLK}
The Color Glass Condensate (CGC) is the effective field theory that governs the IP-Glasma model. It separates soft and hard degrees of freedom whereby the valence partons are treated as external sources for the soft gluons. The JIMWLK renormalization group equations \cite{JalilianMarian:1996xn} integrate out quantum modes around the classical background field, thereby incorporating them into the source term for the small-$x$ gluons. The result is an effective Lagrangian of the same form as the CGC, but with an altered, rapidity dependent, source term. In this study, these effective source terms are not included in Gauss' Law. This will be discussed in more detail in \cite{InPrep}. Numerically, we utilize the Langevin step formulation developed in \cite{Lappi:2012vw} with the modified kernel used in \cite{Schenke:2016ksl}, the first study to implement JIMWLK evolution in the IP-Glasma framework. 
%==========================================%
\section{Fields and Pressure}

\begin{figure}[hb!]
  \centering
  \begin{tabular}{cc}
        \includegraphics[width=0.5\textwidth]{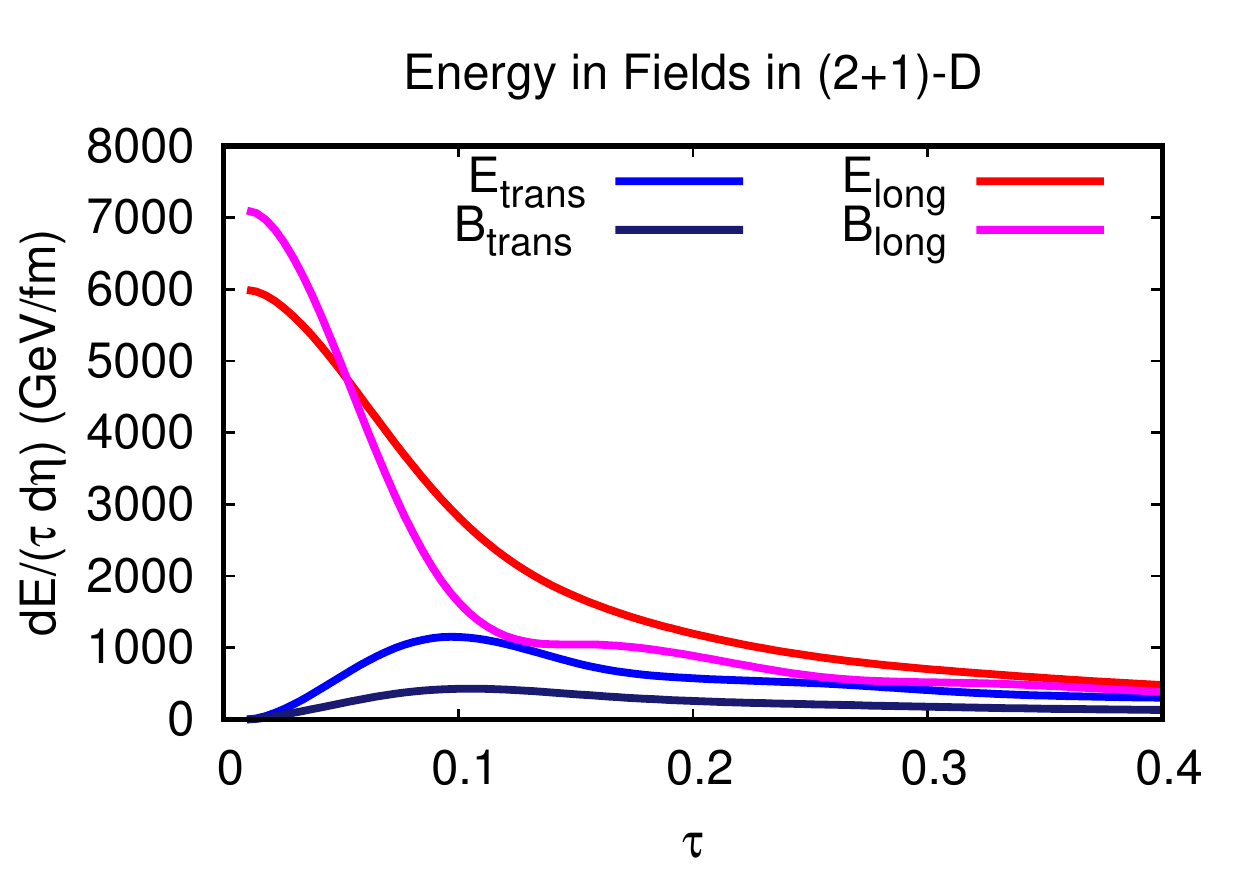}
        \includegraphics[width=0.5\textwidth]{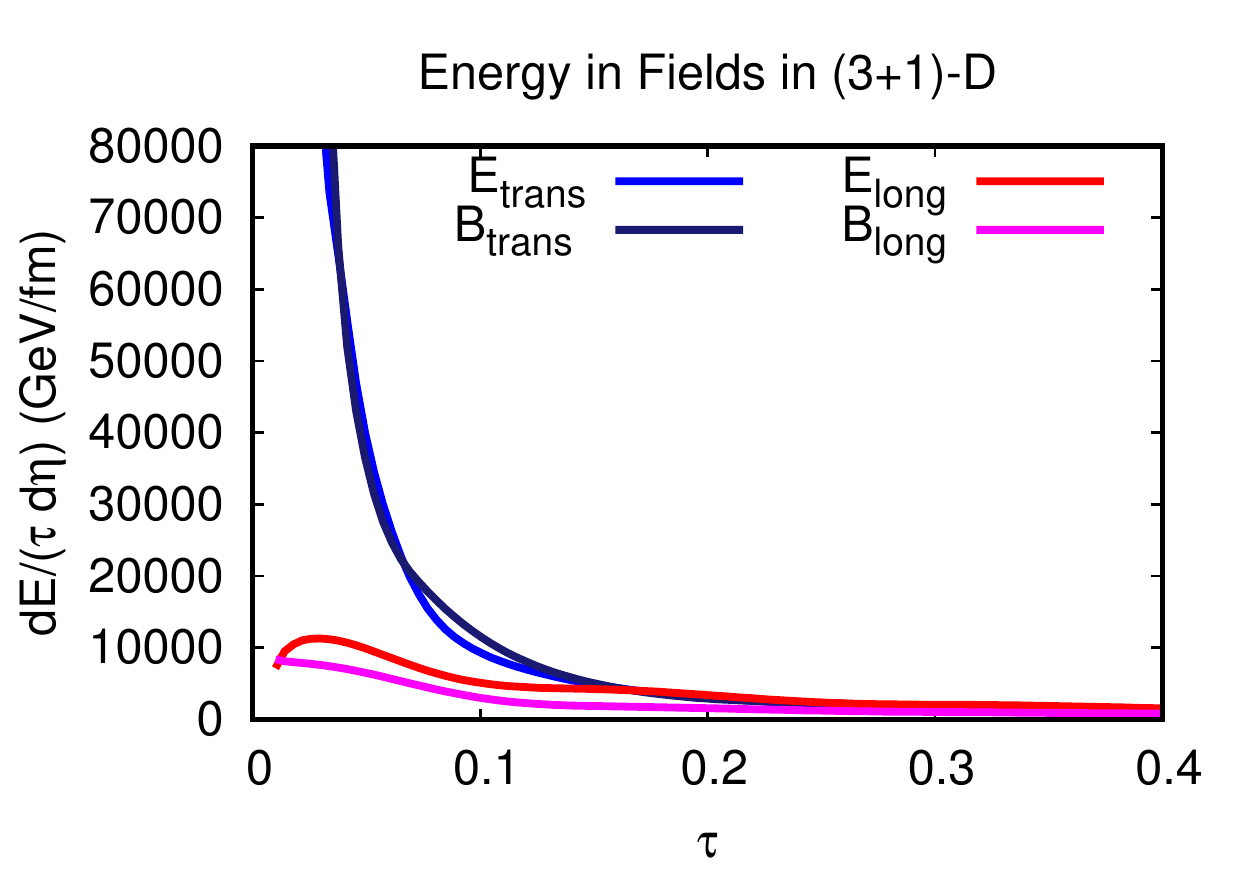}
\end{tabular}
 \caption{Left: The time evolution of the energy density in the different field components in 2+1D. Right: The same quantity as the left panel plotted for the 3+1D implementation. Both results are computed using the same 3+1D software \cite{InPrep}, but with the initial 2+1D and 3+1D setups, respectively.}
 \label{fig1}
\end{figure}

\begin{figure}[hb!]\label{fig:pressure}
\centering
       \includegraphics[width=.5\textwidth]{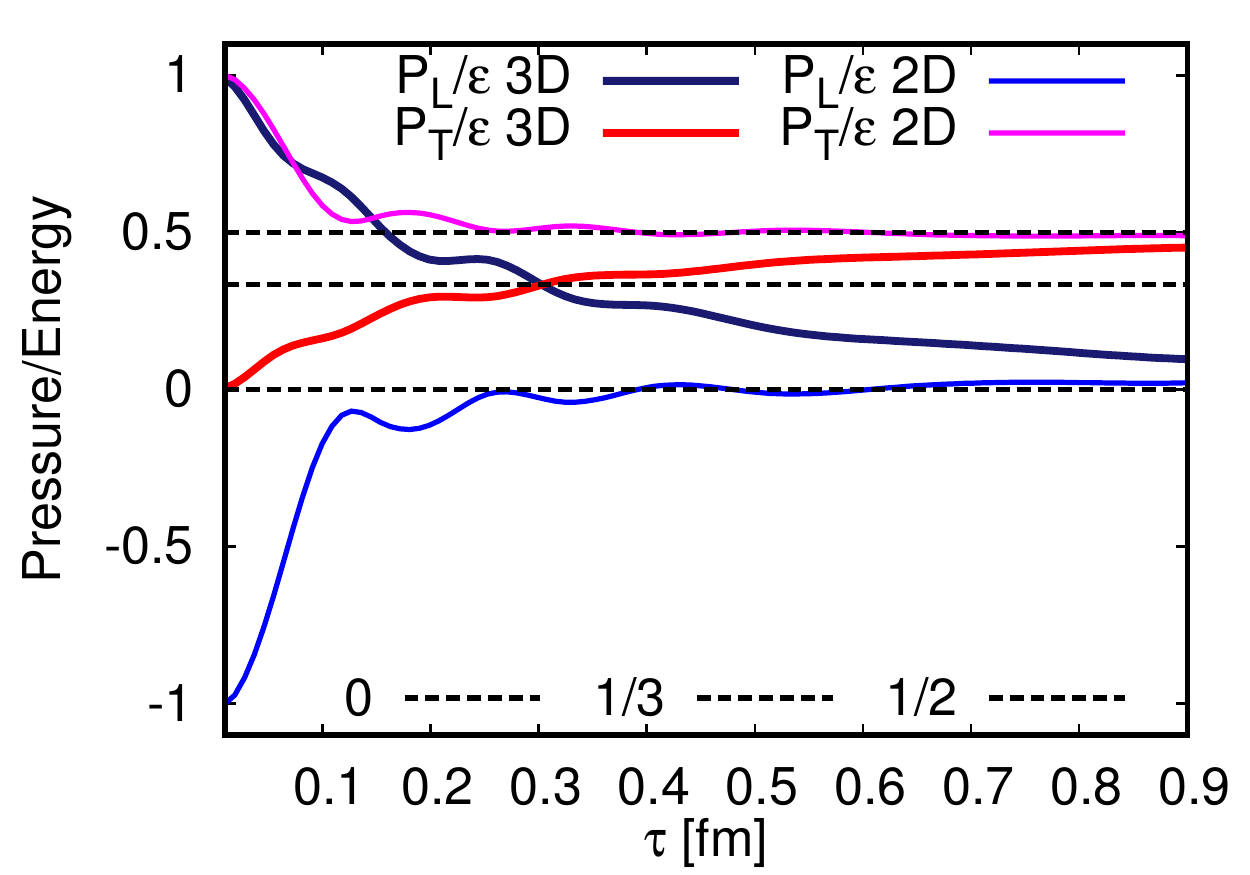}
       \caption{Comparison of the transverse and longitudinal pressures in the 2+1D and 3+1D IP-Glasma formulations. Both results are computed using the same 3+1D software \cite{InPrep}, but with the initial 2+1D and 3+1D setups, respectively.}
       \label{fig2}
\end{figure}

As already discussed, the initial transverse chromo-electric and chromo-magnetic fields vanish in the boost invariant case. In 3+1D, this is no longer true, and in fact these fields have a factor of $1/\tau^2$ in their contribution to the energy density. The energy density in the different fields is given by
%==========================================%
\begin{align}\label{eq:energy_fields}
     \epsilon_{i=x, y} = \frac{1}{2g^2}\frac{1}{\tau^2}\Big [(E^i)^2+(B^i)^2\Big], && \hbox{and} &&
     \epsilon_\eta = \frac{1}{2g^2}\Big [(E^\eta)^2+(B^{\eta})^2\Big].
\end{align}
%==========================================%
This means that instead of vanishing at initial time, as they do in the boost invariant case, the transverse fields actually dominate the energy density at early times, which can be seen clearly in the right panel of fig. [\ref{fig1}]. By typical hydrodynamic initialization times of $\tau=0.2-0.6 \, \rm{fm}$ , the 3+1D fields all have similar contributions to the energy, as is the case in 2+1D.
%==========================================%
%==========================================%
The early time behavior of the fields in 3+1D causes the longitudinal and transverse pressures to behave quite differently than in the boost invariant case. To see why, it is convenient to first express the diagonal components of the stress energy tensor in terms of the quantities defined in eq. [\ref{eq:energy_fields}],
%==========================================%
  \begin{align}
  %\begin{split}
     T^{\tau \tau}&= \epsilon_x+\epsilon_y+\epsilon_{\eta}=\epsilon &&
    T^{ii}=-\, \epsilon_i\,+\epsilon_j+\epsilon_{\eta} \, \biggr \rvert_{\substack{i=x,y\\j \neq i}} &&
     \tau^2 T^{\eta \eta} =\epsilon_x+\epsilon_y\, -\,\epsilon_{\eta}.
  %\end{split}
  \end{align}
%==========================================%
The pressure to energy ratios are given by, 
\begin{align}
      \frac{P_L}{\epsilon} = \frac{\tau^2 T^{\eta\eta}}{T^{\tau \tau}} &&
    \frac{P_\perp}{\epsilon} = \frac{T^{xx}+T^{yy}}{2T^{\tau \tau}}.
  \end{align}
%==========================================%
As can be seen in fig. [\ref{fig:pressure}], the $\tau \xrightarrow{}0^+$ limit is quite different in 2+1D and 3+1D,
%==========================================%
\begin{align}
       \lim_{\tau\to 0^+} \frac{P_L}{\epsilon} =
     \left\{\begin{array}{lr}
            \frac{\epsilon_x+\epsilon_y}{\epsilon_x+ \epsilon_y}=1 & \hbox{in 3+1D}\\
            \frac{-\epsilon_\eta}{\epsilon_\eta}=-1 & \hbox{in 2+1D}
      \end{array}\right. &&
    \lim_{\tau\to 0^+} \frac{P_\perp}{\epsilon} =
     \left\{\begin{array}{lr}
            \frac{\epsilon_\eta}{\epsilon_x+\epsilon_y} = 0 & \hbox{in 3+1D}\\
            \frac{\epsilon_\eta}{\epsilon_\eta}=1 & \hbox{in 2+1D} .
      \end{array}\right. 
\end{align}
%==========================================%
%==========================================%
Due to the tracelessness of $T^{\mu\nu}$, the intersection of the pressures necessarily occurs at $\epsilon/3$, the condition for pressure isotropy. The pressure does not remain isotropic, however, and approaches the 2+1D asymptotic behavior for large $\tau$, as the longitudinal pressure free-streams towards zero in both cases. 
\section{Conclusion}
We have generalized the 2+1D initial condition to include a non-zero $A_\eta$ and a local solution of Gauss' Law. These changes are necessary for an initial condition that is consistent with the non-zero $\eta$-derivatives present in a non-boost-invariant system, and have important consequences. Among these are the early time evolution of the energy density and pressure.   
\section{Acknowledgements}
This work is supported in part by the Natural Sciences and Engineering Research Council of Canada. Computation for this work was done in part on the Compute Canada supercomputers Cedar, maintained by West Grid, and Guillimin, maintained by Calcul Qu\'ebec. SM acknowledges funding from The Fonds de recherche du Qu\'ebec - Nature et technologies (FRQNT) through the Programme de Bourses d'Excellence pour \'Etudiants \'Etrangers (PBEEE). CG is grateful to the Canada Council for the Arts for support through its Killam Research Fellowship Program. 



%% References


\bibliographystyle{elsarticle-num}

\end{document}